\documentclass[prl,a4paper,twocolumn,nofootinbib]{revtex4}
\usepackage[a4paper, hdivide={1.91cm,,1.165cm}, vdivide={2.0cm,,2.33cm}]{geometry}

\usepackage{amstext,amssymb}
\usepackage[intlimits]{amsmath}
\usepackage[english]{babel}
\usepackage[ansinew]{inputenc}
\usepackage{graphicx}
\usepackage{units}
\usepackage{hyperref}
\usepackage{color}

\hypersetup{
    pdftitle={Neutrinoless Quadruple Beta Decay},
    pdfauthor={Julian Heeck, Werner Rodejohann}
}

\begin{document}

\let \Lold \L
\def \L {\mathcal{L}} %Lagrangian density
\newcommand{\hc}{\mathrm{h.c.}}
\newcommand{\re}{\mathrm{Re}\,}
\newcommand{\im}{\mathrm{Im}\,}

\title{Neutrinoless Quadruple Beta Decay}

\author{Julian \surname{Heeck}}
\email[Electronic address: ]{julian.heeck@mpi-hd.mpg.de}
\affiliation{Max-Planck-Institut f\"ur Kernphysik, Saupfercheckweg 1, 69117 Heidelberg, Germany}

\author{Werner \surname{Rodejohann}}
\email[Electronic address: ]{werner.rodejohann@mpi-hd.mpg.de}
\affiliation{Max-Planck-Institut f\"ur Kernphysik, Saupfercheckweg 1, 69117 Heidelberg, Germany}


\begin{abstract}

We point out that lepton number violation is possible even if neutrinos are Dirac particles. We illustrate this by constructing a simple model that allows for lepton number violation by four units only. As a consequence, neutrinoless double beta decay is forbidden, but neutrinoless quadruple beta decay is possible: $(A,Z) \to (A,Z+4) + 4\, e^-$. We identify three candidate isotopes for this decay, the most promising one being ${}^{150}\mathrm{Nd}$ due to its high $Q_{0\nu 4\beta}$-value of $\unit[2]{MeV}$. Analogous processes, such as neutrinoless quadruple electron capture, are also possible. The expected lifetimes are extremely long, and experimental searches are challenging.  

\end{abstract}

\maketitle

%%%%%%%%%%%%%%%%%%%%%%%%%%%%%%%%%%%%%%%%%%%%%%%%%%%

\section{Introduction}
\label{sec:introduction}

Of all the open questions concerning neutrinos---mass scale and hierarchy, 
possible CP violation, origin of the mixing pattern---the conceptually most 
interesting has to be its very nature: is the neutrino its own antiparticle, and hence a 
Majorana fermion, or do neutrino and antineutrino differ, making the neutrino a Dirac 
particle like all the other fermions of the Standard Model (SM). 
The key observation here would be neutrinoless double beta decay 
$(0\nu2\beta)$~\cite{Rodejohann:2012xd}, 
$ (A,Z) \to (A,Z+2) + 2\, e^- $, 
because the observation of this $\Delta L = 2$ process unambiguously confirms the 
Majorana nature of neutrinos~\cite{Schechter:1981bd}. 
Other $\Delta L = 2$ signatures, ranging from low-energy processes like 
neutrinoless double electron capture to collider processes, have also been proposed and tested (a collection of references 
can be found in Refs.~\cite{Atre:2009rg,Rodejohann:2012xd}),  
but all experiments came up empty so far.

The necessary lepton number violation (LNV) by two units, $\Delta L = 2$, can be 
realized directly with a tree level Majorana mass term, or indirectly via 
diagrams containing two vertices with $\Delta L = 1$, one example being $R$-parity 
violating supersymmetry \cite{RPV}. 
However, it is most often overlooked that LNV and Majorana neutrinos are not necessarily 
connected. For instance, there are non-perturbative processes in the SM that violate lepton (and baryon) 
number by three units \cite{sphalerons}, $\Delta L = \Delta B = 3$, which obviously 
do not lead to Majorana neutrinos, and are in fact perfectly compatible with 
Dirac neutrinos. 

In this letter we will entertain the possibility that LNV occurs only by 
four units, and that $\Delta L = 2$ processes are forbidden; neutrinos are then Dirac particles. 
We will realize those lepton number violating Dirac neutrinos in a simple model based on a spontaneously broken
$U(1)_{B-L}$. As an interesting consequence, neutrinoless quadruple beta decay ($0\nu 4\beta$), 
\begin{align}
(A,Z) \to (A,Z+4) + 4\, e^- \, , 
\label{eq:0nbbbb}
\end{align}
is allowed. This novel nuclear decay process plays for our framework the role that neutrinoless double 
beta decay plays for Majorana neutrinos:
it will be the dominant possible LNV process for Dirac neutrinos, which is surely of great 
conceptual interest even if the decay rates that we estimate are tiny.\footnote{One might think that $0\nu 3\beta$ should be the next probable neutrinoless beta decay after $0\nu 2\beta$. This process would however violate Lorentz symmetry, similar to neutrinoless single beta decay $n\rightarrow p + e^-$.}

We will first construct a simple toy model that forbids Majorana neutrinos 
but allows for LNV by four units. Then we will search for interesting isotopes that can undergo $0\nu 4\beta$ and estimate the 
expected lifetimes. Interestingly, all three isotopes that we identify as potential 
$0\nu 4\beta$-emitters ($^{96}$Zr, $^{136}$Xe, and $^{150}$Nd) 
are familiar from searches for neutrinoless double 
beta decay. The most interesting candidate is $^{150}$Nd, with a $Q_{0\nu 4\beta}$-value of $\unit[2.079]{MeV}$. 
We also identify four candidates for neutrinoless quadruple electron capture and related processes ($^{124}$Xe, $^{130}$Ba, $^{148}$Gd, and $^{154}$Dy).
More detailed studies regarding model building aspects, collider phenomenology and cosmological aspects will be presented elsewhere.

\section{Simple Model for \texorpdfstring{$\boldsymbol{\Delta (B-L)=4}$}{Delta (B-L) = 4}}
\label{sec:simple_model}

We introduce three right-handed neutrinos $\nu_R$ (RHNs) to the SM, which results in Dirac masses for 
the neutrinos after spontaneous electroweak symmetry breaking. 
A striking feature of the chiral fermion content of the SM$+\nu_R$ is the existence of a new, 
accidental, anomaly-free symmetry  $U(1)_{B-L}$, which can therefore be consistently gauged in 
addition to the SM gauge group. Breaking $B-L$ by a scalar $\phi$ with charge 
$|B-L| = 4$ can then lead to a remaining discrete symmetry group $\mathbb{Z}_4^L$ in the lepton 
sector, which protects the Dirac structure of neutrinos and still allows for LNV processes.
Quartic LNV operators for Dirac neutrinos were also mentioned in a study of anomaly-free discrete $R$-symmetries in Ref.~\cite{Chen:2012jg}.

For a simple realization of this idea, we work with a gauged $U(1)_{B-L}$ symmetry, 
three RHNs $\nu_R \sim -1$, one scalar $\phi\sim 4$, and one scalar 
$\chi \sim -2$, all of which are SM-singlets. 
The Lagrangian takes the form
\begin{align}
\begin{split}
 \L \ &= \L_\mathrm{SM} +\L_\mathrm{kinetic}(\nu_R,\phi,\chi) +\L_{Z'}- V(H,\phi,\chi) \\
&\quad + \left( y_{\alpha\beta} \overline{L}_\alpha H \nu_{R,\beta} 
+ \kappa_{\alpha\beta} \chi \, \overline{\nu}_{R,\alpha} \nu_{R,\beta}^c  +\hc\right) ,
\end{split}
\end{align}
$H$ being the SM Higgs doublet. The phenomenology of the accompanying $Z'$ boson, 
described in $\L_{Z'}$, is not important here. 
Working in the diagonal charged lepton basis, the neutrinos obtain the Dirac mass matrix 
$M_{\alpha\beta} \equiv |\langle H \rangle| \, y_{\alpha\beta} $ upon electroweak symmetry breaking. 
A bi-unitary transformation can be used to diagonalize this mass matrix via
$U^\dagger M \, V = \mathrm{diag} (m_1,m_2,m_3)$, where $U$ is the lepton mixing matrix 
relevant for electroweak charged-current interactions. 
Contrary to other models with Dirac neutrinos, the right-handed transformation matrix 
$V$ does not drop out, but can be absorbed by the complex symmetric Yukawa coupling 
matrix $\kappa_{\alpha \beta} = \kappa_{\beta \alpha}$, which is non-diagonal in general. 

The scalar potential of our model is of the simple form
\begin{align}
\begin{split}
 &V(H,\phi,\chi) \equiv \sum_{X= H, \phi,\chi} \left(\mu_X^2 |X|^2 + \lambda_X |X|^4\right)\\
 &\quad + \lambda_{H\phi} |H|^2 |\phi|^2 + \lambda_{H\chi} |H|^2 |\chi|^2 
+ \lambda_{\chi\phi} |\chi|^2 |\phi|^2 \\
&\quad - \left( \mu \phi \chi^2 + \hc \right) .
\end{split}
\label{eq:potential}
\end{align}
Here, the coefficients $\mu_j$ and $\lambda_j$ have mass dimension one and zero, respectively. 
Assuming $\mu_H^2,\mu_\phi^2 < 0 < \mu_\chi^2$ and appropriate signs and magnitudes 
of the $\lambda_j$, we can easily construct a potential that is bounded from below and 
breaks $SU(2)_L\times U(1)_Y \times U(1)_{B-L}$ to $U(1)_\mathrm{EM}\times \mathbb{Z}_4^L$. 
In order to forbid Majorana neutrinos, 
it is imperative that $\chi$ does not acquire a vacuum expectation value; 
without the last line in Eq.~\eqref{eq:potential}, 
the necessary condition for this would be
\begin{align}
 m_c^2 \equiv \mu_\chi^2+ \lambda_{H\chi} \langle H\rangle^2 + \lambda_{\chi\phi} \langle \phi\rangle^2 > 0\,,
\end{align}
but the $\mu$ term modifies this condition. To see how, let us first 
note that we can chose $\mu$ and $\langle \phi \rangle$ real and positive w.l.o.g.\ using 
phase and $B-L$ gauge transformations.
The $\mu$ term will then induce a mass splitting between the properly 
normalized real (pseudo)scalar fields $\re(\chi)$ and $\im(\chi)$
\begin{align}
 m_{\re(\chi)}^2 = m_c^2 - 2\mu\langle \phi\rangle\,, &&
 m_{\im(\chi)}^2 = m_c^2 + 2\mu\langle \phi\rangle\,,
\end{align}
so the condition $\langle \chi \rangle =0$ becomes equivalent to 
$m_{\re (\chi)}^2>0$, which can be easily satisfied.  

\begin{figure}[tb]
\setlength{\abovecaptionskip}{-1ex}
	\begin{center}
		\includegraphics[width=0.45\textwidth]{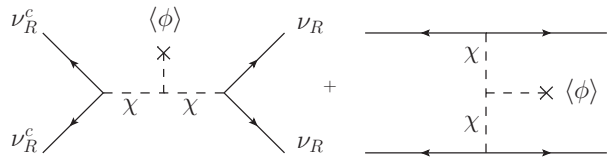}
	\end{center}
		\caption{Tree-level realization of the 
$\Delta L = 4$ operator $(\overline{\nu}^c_R \nu_R)^2$ describing the scattering 
$\nu_R^c \nu_R^c \rightarrow \nu_R \nu_R$.}
	\label{fig:uv_model}
\end{figure}

Neutrinos are hence Dirac particles, but we also obtain effective $\Delta L = 4$ four-neutrino operators by 
integrating out $\chi$ at energies $E\ll m_{\re (\chi)}$, $m_{\im (\chi)}$: 
\begin{align}
\begin{split} 
 \L_\mathrm{eff}^{\Delta L = 4}  &\supset  \frac{1}{2}\left( m_{\im (\chi)}^{-2}- m_{\re (\chi)}^{-2} \right)\left(\kappa_{\alpha\beta}  \overline{\nu}_{R,\alpha} \nu_{R,\beta}^c \right)^2 + \hc ,
\end{split}
\end{align}
see Fig.~\ref{fig:uv_model} for the relevant Feynman diagrams.
For simplicity, we will assume physics at the TeV scale as the source of our 
four-neutrino operators throughout this paper; a discussion of more 
constrained light mediators, as well as of 
other and more complicated models that generate effective four-neutrino operators 
with left-handed neutrinos, will be presented elsewhere. We note that our particular 
example uses a gauged $B-L$ framework; in general however, the observation and the model building 
possibilities that might lead to lepton number violating Dirac neutrinos 
are much broader.

\section{Candidates for \texorpdfstring{$\boldsymbol{0\nu 4\beta}$}{0nu4beta}}
\label{sec:experimental_sigs}

\begin{figure}[b]
\setlength{\abovecaptionskip}{-1ex}
	\begin{center}
		\includegraphics[width=0.45\textwidth]{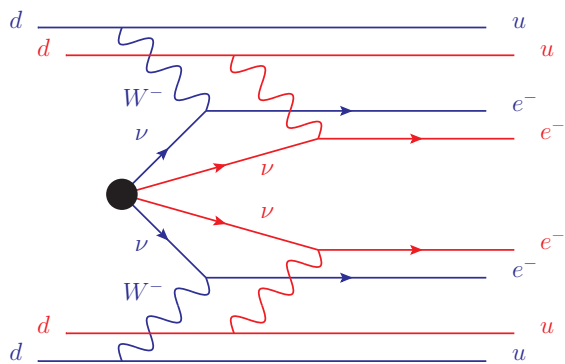}
	\end{center}
		\caption{Neutrinoless quadruple beta decay $4 d \rightarrow 4 u + 4 e^-$ via a 
$\Delta L = 4$ operator $ (\overline{\nu}^c \nu)^2$ (filled circle). Arrows denote flow of 
lepton number, colors are for illustration purposes.}
	\label{fig:quadruple_beta_decay}
\end{figure}

Our model from the last section gave us the effective dimension-six $\Delta L = 4$ operator 
$(\overline{\nu}_R \nu_R^c)^2$, which can lead to an interesting signature in beta decay 
measurements: four nucleons undergo beta decay, emitting four neutrinos; these 
four meet at the effective $\Delta L = 4$ vertex and remain virtual. 
We only see four electrons going out, so at parton level we have 
$4 d \rightarrow 4 u + 4 e^-$, and on hadron level $4 n \rightarrow 4 p + 4 e^-$ 
(Fig.~\ref{fig:quadruple_beta_decay}). Obviously this neutrinoless quadruple beta decay 
($0\nu 4 \beta$) is highly unlikely---more so than $0\nu 2\beta$, 
as it is of fourth order---but one can still perform the exercise of identifying 
candidate isotopes for the decay and estimating the lifetime; constraining the lifetime 
experimentally is of course also possible. 
Besides $0\nu 4 \beta$, one can imagine analogous processes such as 
neutrinoless quadruple electron capture $(0\nu 4 \mathrm{EC})$, 
neutrinoless quadruple positron decay $(0\nu 4\beta^+)$, neutrinoless double electron 
capture double positron decay $(0\nu 2 \mathrm{EC} 2\beta^+)$, etc. 
We will find potential candidates for $0\nu 4 \beta$, $0\nu 2 \mathrm{EC} 2\beta^+$, $0\nu 3 \mathrm{EC} \beta^+$, and $0\nu 4 \mathrm{EC}$.

\begin{figure}[tb]
\setlength{\abovecaptionskip}{-1ex}
	\begin{center}
		\includegraphics[width=0.35\textwidth]{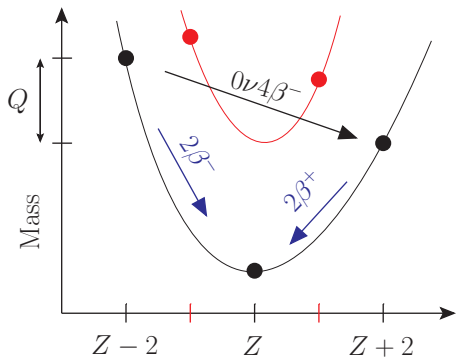}
	\end{center}
		\caption{Three beta-stable even--even nuclei on their mass parabola (black). 
The heaviest isobar $(A, Z-2)$ can decay either via double beta decay into the lowest state 
$(A, Z)$, or via $0\nu 4\beta$ into the medium state $(A, Z+2)$. 
Also shown are the ``forbidden'' odd--odd states in between (red).}
	\label{fig:termschema}
\end{figure}

We will now identify those candidate isotopes for $\Delta L =4$ processes. 
We need to find isotopes which are more stable after the flip $(A,Z) \rightarrow (A, Z\pm 4)$. 
Normal beta decay has to be forbidden in order to handle backgrounds and make the 
mother nucleus sufficiently stable.
Using nuclear data charts~\cite{nuclear_data}, 
we found seven possible candidates: three for $0\nu 4 \beta$, 
four for neutrinoless quadruple electron capture and related decays. They are 
listed in Tab.~\ref{tab:0nu4beta}, together with their $Q$-values, 
competing decay channels, and natural abundance. 
It should be obvious that not all $0\nu 2\beta$ candidates $(A,Z)$ make good $0\nu 4\beta$ candidates, as $(A, Z+4)$ can have a larger mass than $(A,Z)$; it is less obvious that there exist no $0\nu 4\beta$ candidates with beta-unstable daughter nuclei. Using the semi-empirical Bethe--Weizs\"acker mass formula, one can however show that
\begin{align}
 \frac{ M[{}^{A}\!(Z-2)] - M[{}^{A}\!(Z+2)] }{ M[{}^{A}\!(Z-1)] - M[{}^{A}\!(Z+1)] } = 2 \,,
\end{align}
where $M[{}^{A}\!Z] $ denotes the mass of the neutral atom ${}^{A}\!Z$ in its ground state.
Applied to our problem, this means that the mass splitting of the odd--odd states in Fig.~\ref{fig:termschema} (shown in red) is expected to be smaller than the mass splitting of the two $\Delta Z = 4$ nuclei (which is just the $Q$-value, see below), which implies that beta-stable $0\nu 4\beta$ candidates will decay into beta-stable nuclei (this simple argument is confirmed with data charts~\cite{nuclear_data}).

\begin{table}
\renewcommand{\baselinestretch}{1.4}\normalsize
\centering
\begin{tabular}[b]{c|c|c|c}
\hline
\hline
 & $Q_{0\nu 4\beta}$ & Other decays & NA\\
 \hline 
${}^{96}_{40}\mathrm{Zr} \to {}^{96}_{44}\mathrm{Ru}$ & $0.629$ & $\tau_{1/2}^{2\nu 2\beta} \simeq 2 \times 10^{19}$ & $2.8$ \\
${}^{136}_{54}\mathrm{Xe} \to {}^{136}_{58}\mathrm{Ce}$ & $0.044$ &  $\tau_{1/2}^{2\nu 2\beta} \simeq 2 \times 10^{21}$ & $8.9$ \\
${}^{150}_{60}\mathrm{Nd} \to {}^{150}_{64}\mathrm{Gd}$ & $2.079$ &  $\tau_{1/2}^{2\nu 2\beta} \simeq 7 \times 10^{18}$ & $5.6$ \\
\hline 
\hline 
 & $Q_{0\nu 4 \mathrm{EC}}$ &   &   \\
\hline
${}^{124}_{54}\mathrm{Xe} \to {}^{124}_{50}\mathrm{Sn}$ & $0.577$ & --- & $0.095$ \\
${}^{130}_{56}\mathrm{Ba} \to {}^{130}_{52}\mathrm{Te}$ & $0.090$ & $\tau_{1/2}^{2\nu 2\mathrm{EC}} \sim 10^{21}$ & $0.106$ \\
${}^{148}_{64}\mathrm{Gd} \to {}^{148}_{60}\mathrm{Nd}$ & $1.138$ & $\tau_{1/2}^{\alpha} \simeq 75$ & --- \\
${}^{154}_{66}\mathrm{Dy} \to {}^{154}_{62}\mathrm{Sm}$ & $2.063$ & $\tau_{1/2}^{\alpha} \simeq 3\times 10^6$ & --- \\
\hline
\hline
 & $Q_{0\nu 3 \mathrm{EC} \beta^+}$ &   &   \\
\hline
${}^{148}_{64}\mathrm{Gd} \to {}^{148}_{60}\mathrm{Nd}$ & $0.116$ & $\tau_{1/2}^{\alpha} \simeq 75$ & --- \\
${}^{154}_{66}\mathrm{Dy} \to {}^{154}_{62}\mathrm{Sm}$ & $1.041$ & $\tau_{1/2}^{\alpha} \simeq 3\times 10^6$ & --- \\
\hline
\hline
 & $Q_{0\nu 2 \mathrm{EC} 2\beta^+}$ &   &   \\
\hline
${}^{154}_{66}\mathrm{Dy} \to {}^{154}_{62}\mathrm{Sm}$ & $0.019$ & $\tau_{1/2}^{\alpha} \simeq 3\times 10^6$ & --- \\
\hline
\hline
\end{tabular}
\renewcommand{\baselinestretch}{1.0}\normalsize
\caption{\label{tab:0nu4beta}
Candidates for nuclear $\Delta L = 4$ processes neutrinoless quadruple 
beta decay and electron capture, the corresponding $Q$-values in \unit{MeV}, 
competing (observed) decay channels with half-life $\tau_{1/2}^j$ in years, and natural abundance (NA) in percent.}
\end{table}

The $Q$-values in Tab.~\ref{tab:0nu4beta} can be readily calculated in analogy to $0\nu 2\beta$. 
In general, the total kinetic energy of the emitted electrons/positrons in a 
$0\nu n\beta^\mp$ decay,  
\begin{align}
 {}^{A}\!Z \rightarrow {}^{A}\!(Z \pm n) + n\, e^\mp \, , 
\end{align}
is given by the $Q$-value, and can be calculated via
\begin{align}
 Q_{0\nu n\beta^-} &= M[{}^{A}\!Z] - M[{}^{A}\!(Z+n)]\,,\\
 Q_{0\nu n\beta^+} &= M[{}^{A}\!Z] - M[{}^{A}\!(Z-n)] - 2 n \, m_e\,.
\end{align}
The term $- 2 n \, m_e$ in $Q_{0\nu n\beta^+} $ already makes 
$0\nu 2\beta^+$ very rare, but neutrinoless quadruple positron decay 
$0\nu 4 \beta^+$ impossible. 
Electron capture with the emission of up to two positrons 
is however permitted, as the $Q$-value for the EC-process 
\begin{align}
 {}^{A}\!Z +k\, e^- \rightarrow {}^{A}(Z-n) + (n-k)\, e^+
\label{eq:EC}
\end{align}
is given by $Q_{0\nu k\mathrm{EC} (n-k) \beta^+} = Q_{0\nu n\beta^+} + 2 k \, m_e$,
allowing above all for neutrinoless quadruple electron capture $0\nu 4\mathrm{EC}$ 
in four isotopes (Tab.~\ref{tab:0nu4beta}).

\begin{figure}[tb]
\setlength{\abovecaptionskip}{-1ex}
	\begin{center}
		\includegraphics[width=0.45\textwidth]{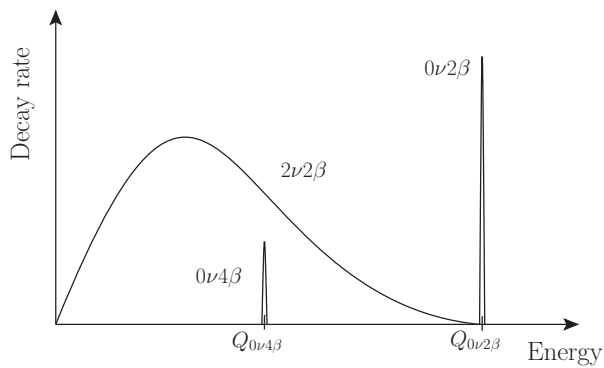}
	\end{center}
		\caption{Sum of kinetic electron energies in the beta decays $0\nu 2\beta$, $2\nu 2\beta$, and $0\nu 4 \beta$. Relative contribution not to scale.}
	\label{fig:energyspectrum}
\end{figure}

Having identified all $\Delta L=4$ 
candidates, we discuss their experimental prospects and challenges in more detail:
Let us first take a look at the most promising element for $0\nu 4\beta$: 
${}^{150}\mathrm{Nd}$. The following decay channels are possible 
(see also Fig.~\ref{fig:termschema}):
\begin{itemize}
\item ${}^{150}_{60}\mathrm{Nd}\rightarrow {}^{150}_{62}\mathrm{Sm}$ via $2\nu 2\beta$, 
i.e.~via the forbidden intermediate odd--odd state ${}^{150}_{61}\mathrm{Pm}$. 
Two neutrinos and two electrons are emitted; the electrons hence have a 
continuous energy spectrum and total energy $E_{e,1}+E_{e,2} < \unit[3.371]{MeV}$. 
This decay has been observed with a half-life of $\unit[7\times 10^{18}]{yrs}$.
\item ${}^{150}_{60}\mathrm{Nd} \rightarrow {}^{150}_{64}\mathrm{Gd}$ via $0\nu 4\beta$. 
Four electrons with continuous energy spectrum and summed energy 
$Q_{0\nu 4\beta} = \unit[2.079]{MeV}$ are emitted. In this special case, the daughter 
nucleus is $\alpha$-unstable with half-life 
$\tau^\alpha_{1/2} ({}^{150}_{64}\mathrm{Gd}\rightarrow {}^{146}_{62}\mathrm{Sm}) 
\simeq \unit[2\times 10^6]{yrs}$.
\end{itemize}
A sketch of the summed electron energy spectrum is shown in 
Fig.~\ref{fig:energyspectrum}.  
$Q_{0\nu 4\beta}$ will always sit somewhere in the middle of the continuous 
spectrum,\footnote{We note that if neutrinos are Majorana particles the decay 
${}^{150}_{60}\mathrm{Nd} \rightarrow {}^{150}_{62}\mathrm{Sm}$ via $0\nu 2\beta$ is possible. 
Two mono-energetic electrons would be emitted with total energy $Q_{0\nu 2\beta} = \unit[3.371]{MeV}$.} 
so one would have to identify the four electrons in order to remove the 
$2\nu 2\beta$ background. 
This still leaves other backgrounds to be considered, e.g.~the scattering of the two $2\nu 2\beta$ electrons off of atomic electrons, which can effectively lead to four emitted electrons (and two neutrinos). Since $Q_{0\nu 4\beta} < Q_{2\nu 2\beta}$, the sum of the electron energies will be continuously distributed and can overlap the discrete $Q_{0\nu 4\beta}$ peak. A dedicated discussion of this and other possible backgrounds goes far beyond the scope of this letter.

As an alternative to direct searches, one could even omit an energy measurement and just look at the 
transmutation ${}^{150}\mathrm{Nd} \rightarrow {}^{150}\mathrm{Gd}$ using, e.g., 
chemical methods; as the background for ${}^{150}\mathrm{Nd} \rightarrow {}^{150}\mathrm{Gd}$ 
is basically nonexistent---the SM-allowed $4\nu 4\beta$ is killed by the $Q$-dependence 
of the eight-particle phase space $G_{4\nu 4\beta}\sim Q^{23}$ 
(compared to the four-particle phase space $G_{0\nu 4\beta}\sim Q^{11}$), 
and $0\nu 2 \beta$ would be seen long before we ever see the double $0\nu 2\beta$ 
that mimics $0\nu 4 \beta$. Hence, this transmutation suffices to test $0\nu 4\beta$. 
In case of ${}^{150}\mathrm{Nd}$, the instability of the daughter 
nucleus ${}^{150}\mathrm{Gd}$ can even be advantageous, as the resulting 
alpha particle provides an additional handle to look for the decay.\footnote{The alpha decay is however too slow to be used in coincidence with $0\nu 4\beta$.} The necessary 
macroscopic number of daughter elements will of course result in 
weak limits compared to dedicated $0\nu 4\beta$ searches in $0\nu 2\beta$ experiments. 
However, for elements not under consideration in $0\nu 2\beta$ experiments, 
this could be a viable and inexpensive way to test $0\nu 4\beta$.

There is also the possibility of decay into an excited state, 
${}^{150}_{60}\mathrm{Nd} \rightarrow {}^{150}_{64}\mathrm{Gd}^*$ via $0\nu 4\beta$.
The excited final state will reduce the effective $Q$-value---by $\unit[0.638]{MeV}$ ($\unit[1.207]{MeV}$) for the lowest $2^+$ ($0^+$) state---and produce detectable photons. 

All the above holds similarly for ${}^{96}\mathrm{Zr}$ and ${}^{136}\mathrm{Xe}$ as well. 
Both have much smaller $Q$-values---which theoretically reduces the rate---but 
$\alpha$-stable daughter nuclei. The non-solid structure of 
xenon makes it in principle easier to check for the transmutation into cerium; 
furthermore, the EXO~\cite{Ackerman:2011gz} $0\nu 2\beta$ experiment is currently 
running and could check for $0\nu 4\beta$, should their detector be 
sensitive at these energies and not flooded by backgrounds.
${}^{96}\mathrm{Zr}$ is a better candidate due to its higher $Q$-value, but there 
are no dedicated ${}^{96}\mathrm{Zr}$ experiments planned. 
Still, the NEMO collaboration could set limits on ${}^{96}\mathrm{Zr} 
\xrightarrow{0.629}  {}^{96}\mathrm{Ru}$ by reanalyzing their data from Ref.~\cite{Zr96}. 
${}^{150}\mathrm{Nd}$ is by far the best candidate, due to the high $Q_{0\nu 4\beta} $-value. 
Coincidentally, it also has a high $Q_{0\nu 2\beta}$-value, which makes it a popular 
isotope to test for $0\nu 2\beta$, with some existing and planned 
experiments~\cite{Rodejohann:2012xd}. Once again, NEMO might already 
be able to constrain ${}^{150}\mathrm{Nd} \xrightarrow{2.079}  {}^{150}\mathrm{Gd}$ 
with their data~\cite{Nd150}.

The $0\nu 4 \mathrm{EC}$ channels in Tab.~\ref{tab:0nu4beta} 
lead to a similar transmutation behavior as discussed above for $0\nu 4\beta^-$, 
and can be checked in the same way. Note that the energy-gain $Q_{0\nu 4\mathrm{EC}}$ will here be carried away by photons instead of electrons; the captured electrons will be taken out of the $K$ and $L$ shells, resulting in a cascade of X-ray photons.
The $Q$-values of ${}^{148}\mathrm{Gd}$ and ${}^{154}\mathrm{Dy}$ are high enough to also undergo $0\nu 3\mathrm{EC} \beta^+$; ${}^{154}\mathrm{Dy}$ is the only isotope capable of $0\nu 2\mathrm{EC} 2\beta^+$.
This can give rise to distinguishable signatures due to the additional $\unit[511]{keV}$ photons from electron--positron annihilation. 
The comparatively fast $\alpha$-decay of ${}^{148}\mathrm{Gd}$ and ${}^{154}\mathrm{Dy}$---and the 
fact that they have to be synthesized from scratch---make them however very challenging 
probes for $\Delta L = 4$, despite their large $Q$-values. 
${}^{124}\mathrm{Xe}$ might then be the best element to test for 
$0\nu 4 \mathrm{EC}$; unfortunately, the enriched xenon used by EXO contains 
almost no ${}^{124}\mathrm{Xe}$, so $0\nu 4 \mathrm{EC}$ is currently hard to test 
(dark matter experiments using xenon can in principle be used, 
as they contain ${}^{124}\mathrm{Xe}$). Resonant enhancement of the $0\nu 4 \mathrm{EC}$ rates, 
as discussed for the $0\nu 2 \mathrm{EC}$ mode \cite{ECEC}, might boost the signal.

Apparently, $\Delta L = 4$ signals are in general easier to test via the $0\nu 4\beta$ channels, 
with both ${}^{96}\mathrm{Zr}$ and ${}^{150}\mathrm{Nd}$ as more favorable 
isotopes when it comes to $Q$-values and natural abundance.

\section{Rates for \texorpdfstring{$\boldsymbol{0\nu 4\beta}$}{0nu4beta}}
\label{sec:rates}

Let us estimate some rates. 
Similar to $0\nu 2\beta$, the half-life of $0\nu 4\beta$ can approximately be factorized as 
\begin{align}
 \left[ \tau_{1/2}^{0\nu 4\beta}\right]^{-1} = G_{0\nu 4\beta} |\mathcal{M}_{0\nu 4\beta}|^2 \, ,
\end{align}
where $G_{0\nu 4\beta}$ denotes the phase space and $\mathcal{M}_{0\nu 4\beta}$ the nuclear transition 
matrix element (including the particle physics parameters) 
facilitating the process. Using an effective 
$\Delta L = 4$ vertex $(\overline{\nu}_L \nu_L^c)^2/\Lambda^2$ gives 
$\mathcal{M}_{0\nu 4\beta} \propto G_F^4/p_\nu^4 \Lambda^2$, just by counting propagators. 
For the virtual neutrino momentum $p_\nu$ we will use the inverse distance between the 
decaying nucleons, $p_\nu \sim |q| \sim \unit[1]{fm^{-1}} \simeq \unit[100]{MeV}$. 
The phase-space factor for the four final particles is the same as the one in 
$2\nu 2\beta$ (proportional to $Q^{11}$ for $Q\gg m_e$~\cite{Suhonen:1998ck}), which also tells us that each of the four electrons 
will be distributed just like the electrons in $2\nu 2\beta$, with a 
different $Q$-value, of course.
Purely on dimensional grounds we can then estimate the dependence of the half-life on our parameters as
\begin{align}
 \left[ \tau_{1/2}^{0\nu 4\beta}\right]^{-1} \propto Q^{11} \left( \frac{G_F^4}{q^4 \Lambda^2} \right)^2 \, q^{18} \,,
\end{align}
where the last factor is included to obtain the correct mass dimension.
The above estimate is only valid for large $Q$-values, as it assumes massless electrons; 
the low $Q_{0\nu 4\beta}$ of most elements in Tab.~\ref{tab:0nu4beta} render (some of) the four 
electrons non-relativistic and make necessary a more accurate calculation of the phase space. 
To partially cancel the uncertainties, we can approximate that 
the phase space for $0\nu 4\beta$ and $2\nu 2\beta$ is overall similar and 
consider the ratio (for ${}^{150}\mathrm{Nd}$ and $|q|\simeq \unit[100]{MeV}$)
\begin{align}
 \frac{\tau_{1/2}^{0\nu 4\beta}}{\tau_{1/2}^{2\nu 2\beta}} 
\simeq \left(\frac{Q_{0\nu 2\beta}}{Q_{0\nu 4\beta}} \right)^{11} \left(\frac{\Lambda^4}{q^{12} G_F^4} \right)
\simeq  10^{46}\, \left(\frac{\Lambda}{\unit{TeV}}\right)^4 .
\end{align}
This is of course a rough estimate, and a better calculation, dropping the 
implicitly used closure approximation, including effects of 
the nuclear Coulomb field etc., will certainly change this rate. 
To this effect we stress a difference between $0\nu 2\beta$ and $0\nu 4\beta$: while the 
former decay proceeds via a kinematically forbidden intermediate state, the latter also 
features an energetically preferred intermediate state $X$, only to rush past it on the 
mass parabola (see Fig.~\ref{fig:termschema}). Since excited states of $X$ can still have a 
lower mass than our initial nucleus, the summation over all these states is important and 
cannot be approximated away as easily as the excited states of an already forbidden 
intermediate state.

Finally, in our simple model from above, we generate the $\Delta L = 4$ operator with RHNs, $(\overline{\nu}_R \nu_R^c)^2$, so each of the neutrinos in Fig.~\ref{fig:quadruple_beta_decay} requires a mass-flip in order to 
couple to the $W$ bosons. The particle physics amplitude is therefore further suppressed 
by a factor $(m_\nu/q)^4 \simeq 10^{-37}$, making this process all the more unlikely.
These mass-flips can be avoided in left--right-symmetric extensions of our model, at the price of replacing the four $W$ bosons in Fig.~\ref{fig:quadruple_beta_decay} with their heavier $W_R$ counterparts.

Even with all our approximations leading to the above estimates,  
one can safely conclude that the half-life for neutrinoless quadruple beta decay is very large, 
at least if physics at the TeV scale is behind it in any way. This may be a too 
conservative approach, because four-neutrino interactions do not suffer from such stringent 
constraints as other four-fermion interactions~\cite{Bilenky:1999dn}. The effective LNV operator $(\overline{\nu}_{L} \nu_{L}^c)^2/\Lambda$ discussed here has not been constrained so far, and the contribution to the well-measured invisible $Z$-width via $Z\to 4\nu$ only gives $\Lambda > 1/(\mathcal{O}(10)\sqrt{G_F}) \sim \unit[20]{GeV}$. This, of course, only holds if the mediator is heavy enough to be integrated out in the first place. Light mediators can significantly increase the rate; the life-time will be minimal if the exchanged particles have masses of the order of $|q| \simeq \unit[100]{MeV}$. For neutrinoless 
double beta decay the gain factor for the half-life is about $10^{16}$~\cite{max}, and we can 
expect something similar here. Given that we have four neutrino propagators, the rate might 
be enhanced by a sizable factor, and therefore experimental searches for $0\nu 4\beta$ 
should be pursued.

While the expected rates for $0\nu 4\beta$ in our proof-of-principle model are unobservably small, more elaborate models---invoking resonances---might overcome this obstacle. Most importantly, the experimental and nuclear-physics aspects of $0\nu 4\beta$ are completely independent of the underlying mechanism, and can therefore be readily investigated.

\section{Conclusion}
\label{sec:conclusion}

Contrary to popular belief, Majorana neutrinos are not a prerequisite for lepton number 
violation, and we have given a simple counterexample of lepton number violating 
Dirac neutrinos in this work. This gives rise to previously undiscussed $\Delta L = 4$ 
processes, the most striking of which would be neutrinoless quadruple beta decay, which can in principle be observable in three nuclei. The most promising 
isotope is ${}^{150}\mathrm{Nd}$ due to its high $Q_{0\nu 4\beta}$-value and natural 
abundance (see Tab.~\ref{tab:0nu4beta}), and existing experiments could 
already be used to test $0\nu 4\beta$. 

Let us stress that the decay should be constrained experimentally, as our theoretical estimates for 
TeV-scale physics induced $0\nu 4\beta$ might be too conservative. 
Not only is it a novel possible decay channel on the nuclear physics side, 
but it contains very interesting conceptual information about the fate 
of the classically conserved lepton number symmetry.

\begin{acknowledgments}
The authors thank Kai Zuber for discussions and comments.
J.H.~thanks Sebastian Lindemann for experimental insights and 
acknowledges support by the IMPRS-PTFS. 
This work was supported by the Max Planck Society in the project MANITOP. 
\end{acknowledgments}

%%%%%%%%%%%%%%%%%%%%%%%%%%%%%%%%%%%%%%%%%%%%%%%%%%%

\end{document}